\documentclass{iau} 
\usepackage{graphicx}
\usepackage{url}
\usepackage{journal-macros}
\usepackage{natbib}

\title[The DR14 APOGEE-TGAS catalogue] 
{The DR14 APOGEE-TGAS catalogue:\\ Precise chemo-kinematics in the extended solar vicinity}

\author[F. Anders et al.]{Friedrich Anders$^{1}$, Anna B. Queiroz$^{3,2}$, Cristina Chiappini$^{1,2}$,\\ Bas{\'i}lio X. Santiago$^{3,2}$, J. G. Fern\'andez-Trincado$^{4}$,
Andres Meza$^{5}$ \and the SDSS-IV/APOGEE Collaboration
 }

\affiliation{
 $^1$Leibniz-Institut f\"ur Astrophysik Potsdam (AIP), An der Sternwarte 16, 14482 Potsdam, Germany; email: {\tt fanders@aip.de} \\[\affilskip]
$^2$Laborat\'orio Interinstitucional de e-Astronomia, - LIneA, Rua Gal. Jos\'e Cristino 77, Rio de Janeiro, RJ - 20921-400, Brazil \\[\affilskip]
$^3$Instituto de F\'\i sica, Universidade Federal do Rio Grande do Sul, Caixa 
Postal 15051, Porto Alegre, RS - 91501-970, Brazil \\[\affilskip]
$^4$Departamento de Astronom{\'i}a, Universidad de Concepci\'on, Casilla 160-C,
Concepci\'on, Chile  \\[\affilskip]
$^5$Facultad de Ingenier\'ia, Universidad Aut\'onoma de Chile, Pedro de Valdivia 425, Santiago, Chile 
}
\pubyear{2017}
\volume{334}  
\jname{Rediscovering our Galaxy}
\editors{C. Chiappini, I. Minchev, E. Starkenburg \& M. Valentini, eds.}

\begin{document}

\maketitle

\begin{abstract}
We describe the DR14 APOGEE-TGAS catalogue, a new SDSS value-added catalogue that provides precise astrophysical parameters, chemical abundances, astro-spectro-photometric distances and extinctions, as well as orbital parameters for $\sim 30,000$ APOGEE-TGAS stars, among them $\sim5,000$ high-quality giant stars within 1 kpc.
\end{abstract}

\firstsection 
\section{Catalogue overview}

\begin{figure*}\centering
 \includegraphics[width=0.9\textwidth]{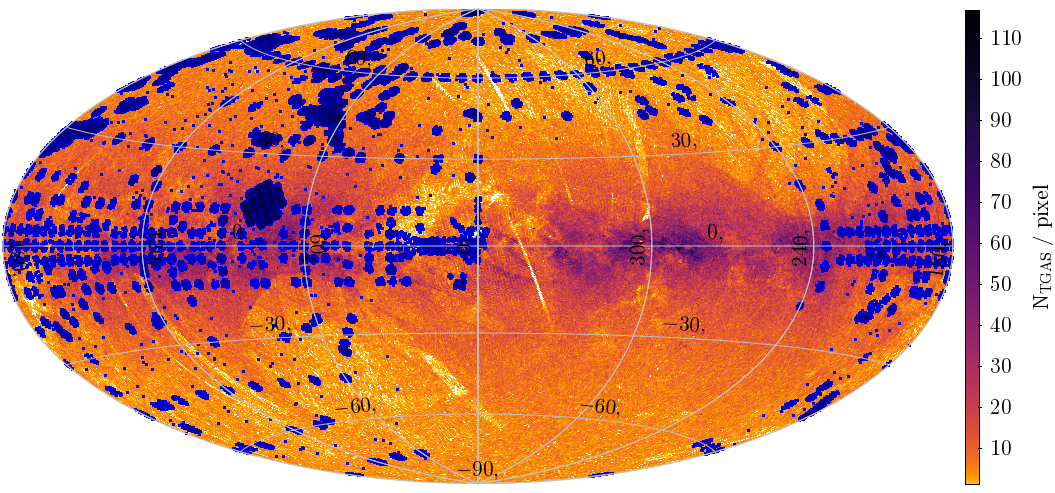}
\caption{Footprint of the DR14 APOGEE-TGAS sample, superimposed on the TGAS source density sky map. The tile size of the HealPix map is 0.21 deg$^2$.}
\label{sky}
\end{figure*}

The first data release of the {\it Gaia} mission \citep{GaiaCollaboration2016a} contains improved parallaxes and proper motions for more than 2 million stars contained in the Tycho-2 catalogue \citep{Hog2000}, among them $40,250$ stars contained in the APOGEE DR14 catalogue. The combined dataset (see Table 1) presents an ideal testbench for chemo-kinematical tagging studies beyond the {\it Hipparcos} volume. 

The Apache Point Observatory Galactic Evolution Experiment (APOGEE; \citealt{Majewski2017}) delivers high-resolution ($R\sim 22,500$) high signal-to-noise ($S/N\sim 100$ pixel$^{-1}$) spectra of primarily red giant stars in the $H$ band ($\lambda=1.51-1.69 \mu$m), enabling the determination of precise ($\sim100$ m/s) radial velocities as well as stellar parameters and chemical abundances of more than 15 elements. For this paper, we use the results from the APOGEE Stellar Parameters and Chemical Abundances Pipeline (ASPCAP; \citealt{GarciaPerez2016}) contained in the Sloan Digital Sky Survey's Fourteenth data release (DR14; \citealt{Abolfathi2017}), together with the recommended stellar parameter cuts and post-calibrations for effective temperature and surface gravity\footnote{\url{http://www.sdss.org/dr14/irspec/parameters/}}. 

The DR14 APOGEE-TGAS catalogue is available as an SDSS-IV DR14 value-added catalogue (VAC). The data can be downloaded as a FITS table from \url{https://data.sdss.org/sas/dr14/apogee/vac/apogee-tgas/apogee_tgas-DR14.fits}. 

\section{Cross-match, distances and extinctions}\label{dists}

We cross-matched the APOGEE DR14 ASPCAP summary file with the {\it Gaia} DR1/TGAS catalogue \citep[][see Table 1 for details]{Lindegren2016}. 
For the stars with measured ASPCAP atmospheric parameters we computed ages, masses, distances, and extinctions using the combined astro-spectro-photometric information and the new Bayesian isochrone-fitting code {\tt StarHorse} (\citealt{Santiago2016, Queiroz2017}). The code computes the posterior probability over the PARSEC 1.2S grid of stellar models \citep{Bressan2012, Tang2014, Chen2014}, taking into account spectroscopic measurements of effective temperature, surface gravity, and global metallicity, as well as multi-band photometry and the parallax measurements from TGAS. All uncertainties were modelled to be Gaussian, and 
the TGAS parallaxes were corrected for systematics as suggested by \citet{Arenou2017}. Our priors are an overall Galactic stellar density prior taking into account thin and thick disc, halo and bulge, 
and a \cite{Chabrier2003} initial-mass function. Tests on a APOGEE-TGAS mock sample of simulated stars showed that our code delivers accurate distance and extinction estimates. The median precision of the reported distances and extinctions for giants amounts to $\sim10\%$ and $\sim0.09$ mag, respectively. For details we refer to \citet{Queiroz2017}.

\begin{table}
  \centering
\caption{Sizes of various useful subsamples of the APOGEE-TGAS sample.}
{\scriptsize
\begin{tabular}{l l cc}
\hline
Name & Requirements & Objects \\
\hline
DR14 APOGEE-TGAS sample & Best 5'' match between allStar.l31-c2.fits and TGAS &  46,033 &\\
Unique DR14 APOGEE-TGAS stars & Internal APOGEE\_ID match &  40,250 &\\
\qquad with measured $T_{\rm eff}$, [Fe/H], $[\alpha$/Fe] & ASPCAP converged & 30,076 &\\
\qquad with ages, distances, and orbits & {\tt StarHorse} \citep{Santiago2016, Queiroz2017} converged & 29,661 &\\
\qquad with most reliable abundances & SNREV$>100$, 4000 K$<T_{\rm eff}<5000$ K, $1<\log g<3.8$, \\ & $\chi^2_{\rm ASPCAP}<10$ no cluster or commissioning stars, \\ 
 & no suspect broad lines or RV combination & 10,499 &\\
\hline
Extended solar-neighbourhood sample & $d<1$ kpc \& most reliable abundances & 4,844 & \\
\qquad blurring-cleaned & 7 kpc $<R_{\rm mean}<9$ kpc, $Z_{\rm max}<1$ kpc & 2,988 &\\
\hline
\end{tabular}
}
 \label{samplesummary}
\end{table}

\section{Orbital parameters}
\label{orbits}

From the full phase-space information ($\alpha, \delta,d,\mu_{\mathrm{\alpha}},\mu_{\mathrm{\delta}},v_{\mathrm{los}}$), the stellar orbits for our sample were calculated 
in a non-axisymmetric Galactic potential that includes a 3D model for the bar
(Model 4 in \citealt{Fernandez-Trincado2016}) 
using the {\tt GravPot16} code\footnote{\url{https://fernandez-trincado.github.io/GravPot16/index.html}}. For $R_{\rm Gal} > 4.5$ kpc, the model was scaled to the observed rotation curve given by \citet{Sofue2015} and $v_{\phi, \rm LSR}=239$ km/s at the solar position ($R_{\rm Gal, \odot}=8.3$ kpc; e.g. \citealt{Bland-Hawthorn2016}). 
From the integrated Galactic orbits, we computed characterizing orbital quantities such as 
$e, R_{\mathrm{mean}}$, and $Z_{\mathrm{max}}$, along with their uncertainties, using a Monte-Carlo technique (e.g. \citealt{Anders2014}).

\begin{figure}\centering
\centering
 \includegraphics[width=0.7\textwidth]{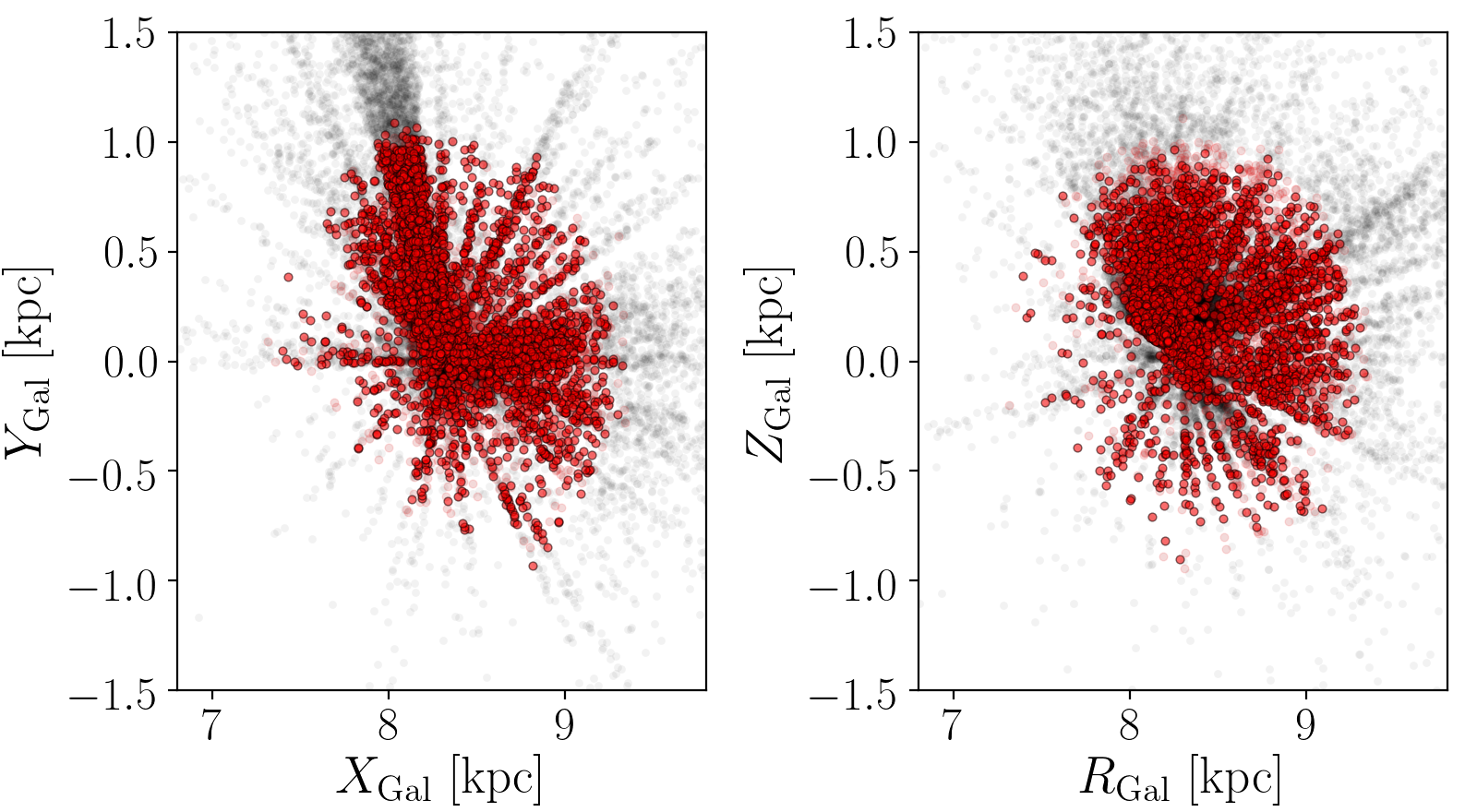}
\caption{Location of the full DR14 APOGEE-TGAS sample (grey) and the extended solar-vicinity sample (red) in Galactocentric coordinates.}
\label{orbit_errors}
\end{figure}

\section{First application: t-SNE dissection of the solar-vicinity chemical-abundance space}

As a first scientific application, we tested for the first time a new non-linear projection method, t-SNE \citep{vanderMaaten2008}, to dissect the local chemical-abundance space. In an accompanying paper (Anders et al., subm.), we show that this method is extremely efficient for finding groups and outliers in multi-dimensional chemical-abundance space, using local high-resolution optical spectroscopic survey data. A subsequent analysis by \citet{Kos2017}, using GALAH data, has shown that it is also possible to reconcile physical star clusters with this method. 

\begin{figure*}\centering
\centering
 \includegraphics[width=0.32\textwidth]{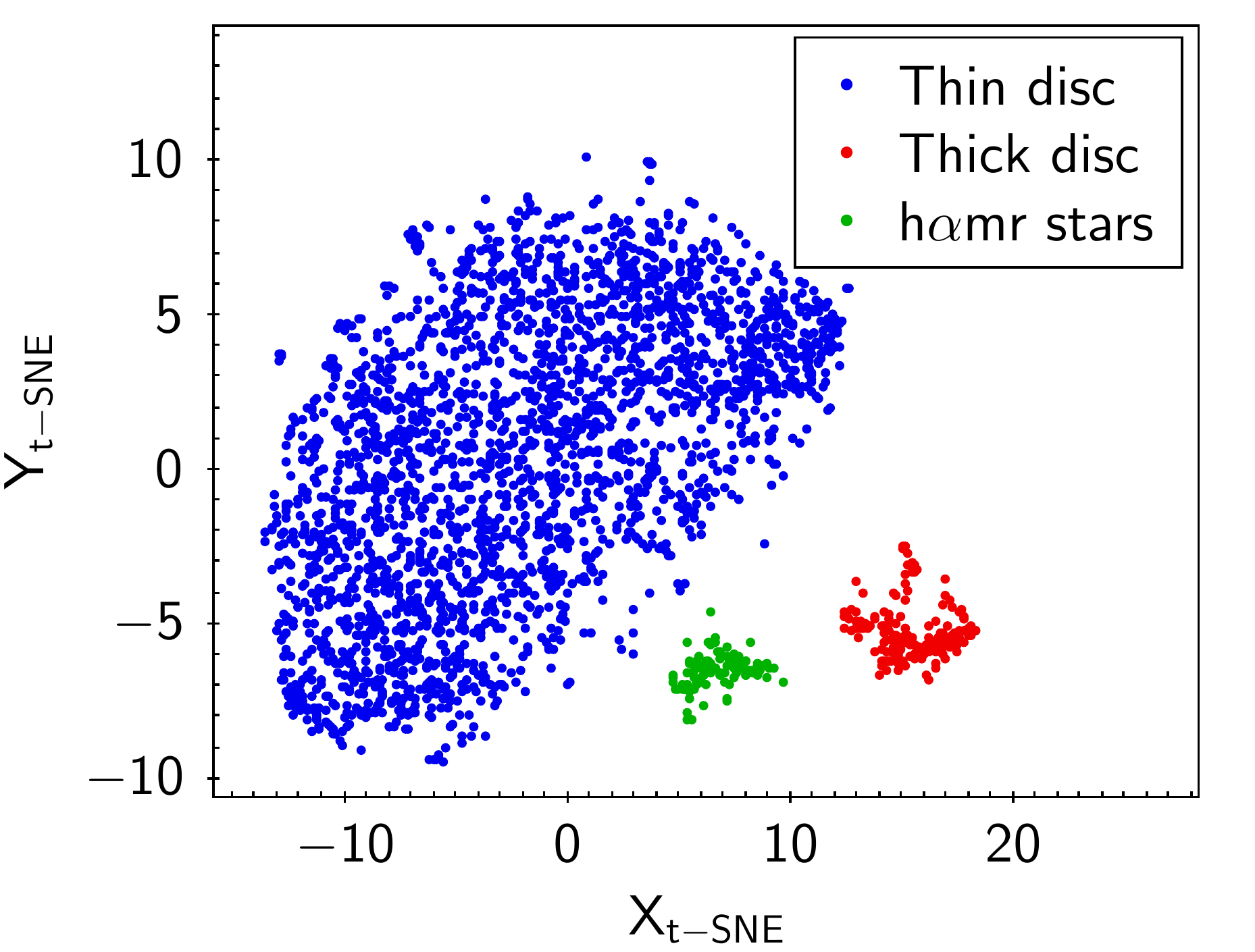}
 \includegraphics[width=0.32\textwidth]{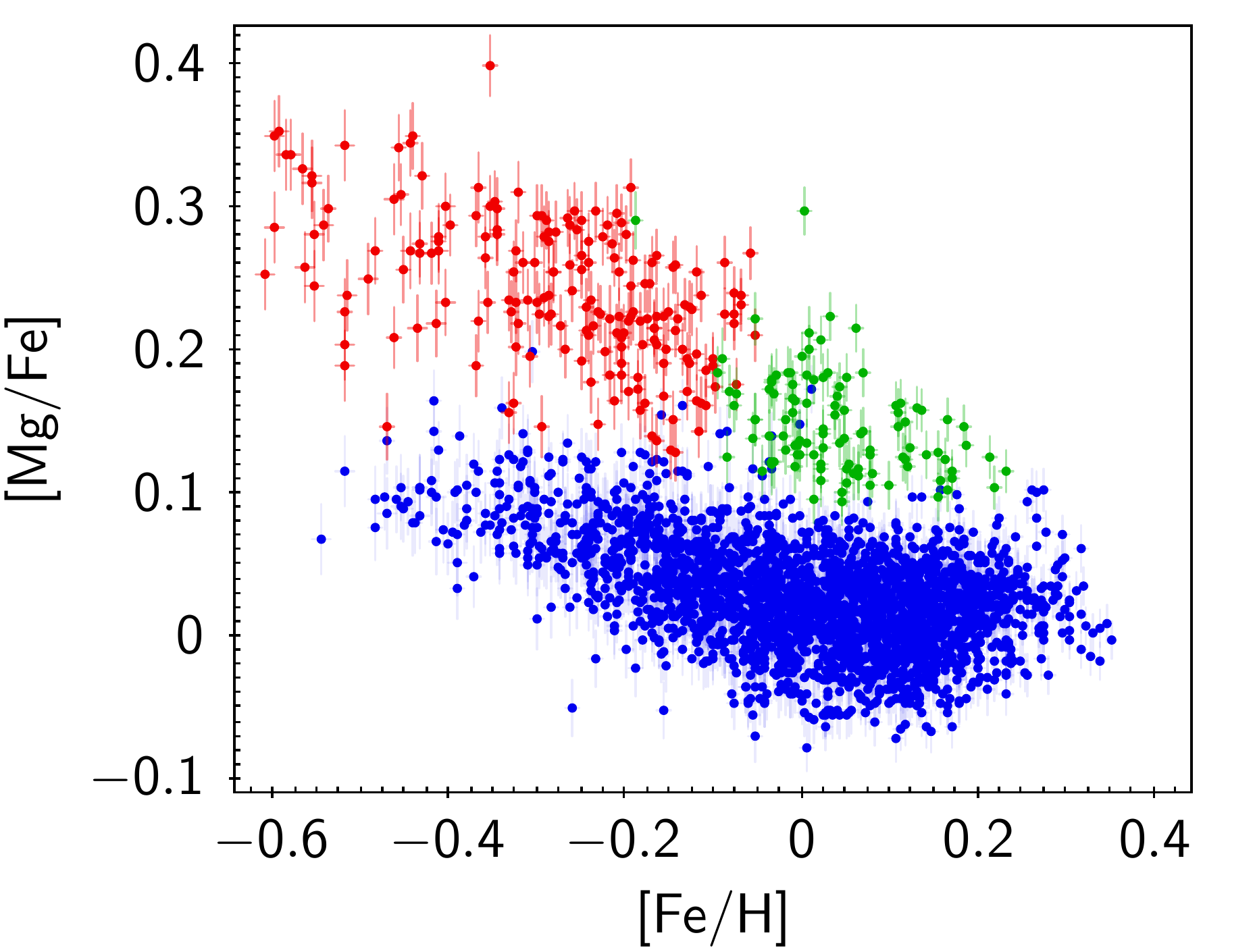}
 \includegraphics[width=0.3\textwidth]{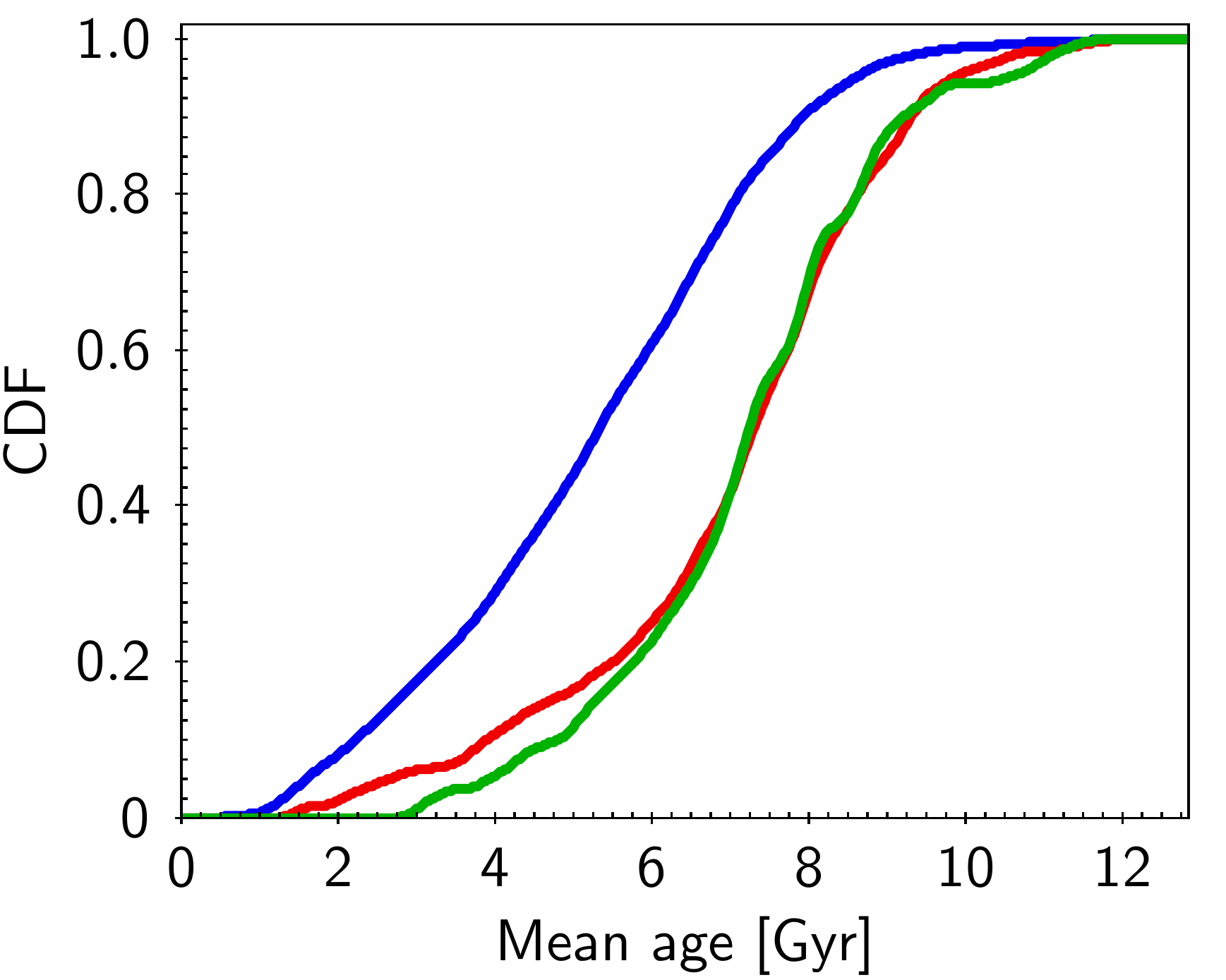}
\caption{Broad t-SNE classification of the blurring-cleaned APOGEE-TGAS extended solar-vicinity sample. Left: Resulting t-SNE projection in 2D, using a perplexity of $p=75$ and the t-SNE hyper-parameters recommended by \citet{vanderMaaten2008}. Middle: [Mg/Fe] vs. [Fe/H] digram, colour-coded by t-SNE population. Right: Cumulative mean age distributions (from {\tt StarHorse}) for the three populations.}
\label{tsne}
\end{figure*}

Fig. \ref{tsne} summarises the results of our preliminary t-SNE analysis of the ASPCAP chemical-abundance space for the blurring cleaned APOGEE-TGAS solar-vicinity sample. The t-SNE map clearly reveals three distinct groups that correspond to the well-known chemical thin and thick discs (blue and red), and the high-[$\alpha$/Fe] metal-rich population (h$\alpha$mr, green; e.g. \citealt{Adibekyan2011}). The two [$\alpha$/Fe]-enhanced groups seem to have very similar age distributions, which suggests that the thick disc and the h$\alpha$mr stars were formed on similar time scales, but in different places. A comprehensive analysis will be presented in a forthcoming paper.

\bibliographystyle{aa}
\bibliography{FA_library}

\begin{acknowledgements}
\begin{scriptsize}
Funding for the Sloan Digital Sky Survey IV has been provided by
the Alfred P. Sloan Foundation, the U.S. Department of Energy Office of
Science, and the Participating Institutions. SDSS-IV acknowledges
support and resources from the Center for High-Performance Computing at
the University of Utah. The SDSS web site is \url{www.sdss.org}.
SDSS-IV is managed by the Astrophysical Research Consortium for the 
Participating Institutions of the SDSS Collaboration including the 
Brazilian Participation Group, the Carnegie Institution for Science, 
Carnegie Mellon University, the Chilean Participation Group, the French Participation Group, Harvard-Smithsonian Center for Astrophysics, 
Instituto de Astrof\'isica de Canarias, The Johns Hopkins University, 
Kavli Institute for the Physics and Mathematics of the Universe (IPMU) / 
University of Tokyo, Lawrence Berkeley National Laboratory, 
Leibniz Institut f\"ur Astrophysik Potsdam (AIP),  
Max-Planck-Institut f\"ur Astronomie (MPIA Heidelberg), 
Max-Planck-Institut f\"ur Astrophysik (MPA Garching), 
Max-Planck-Institut f\"ur Extraterrestrische Physik (MPE), 
National Astronomical Observatory of China, New Mexico State University, 
New York University, University of Notre Dame, 
Observat\'ario Nacional / MCTI, The Ohio State University, 
Pennsylvania State University, Shanghai Astronomical Observatory, 
United Kingdom Participation Group,
Universidad Nacional Aut\'onoma de M\'exico, University of Arizona, 
University of Colorado Boulder, University of Oxford, University of Portsmouth, 
University of Utah, University of Virginia, University of Washington, University of Wisconsin, 
Vanderbilt University, and Yale University.

This work has made use of data from the European Space Agency (ESA)
mission {\it Gaia} (\url{http://www.cosmos.esa.int/gaia}), processed by
the {\it Gaia} Data Processing and Analysis Consortium (DPAC,
\url{http://www.cosmos.esa.int/web/gaia/dpac/consortium}). Funding
for the DPAC has been provided by national institutions, in particular
the institutions participating in the {\it Gaia} Multilateral Agreement.
\end{scriptsize}
\end{acknowledgements}

%
%
%
%
\end{document}